\documentclass[12pt]{article}
\begin{document}
\baselineskip .3in

\begin{center}
{\large{\bf The masses of $P_{c}^{*}(4380)$ and $P_{c}^{*}(4450)$ in the quasi particle diquark model  }} \vskip.2in
R.Ghosh$^1$, A. Bhattacharya $^2$, B.Chakrabarti$^*$\\

\vskip .05in

{\small$^{1,2}$ {Department of Physics, Jadavpur University
Kolkata-700032, India.\\$^{*}$Department of Physics, Jogamaya Devi Collage
Kolkata, India.\\
E-mail: $^{2}$pampa@phys.jdvu.ac.in, $^{*}$$ballari_{-}$chakrabarti@yahoo.co.in }}

\end{center}

\vskip .4in

{\small{The masses of the recently reported by LHCb two
pentaquark charmonium states $P_{c}^{*}(4380)$ and $P_{c}^{*}(4450)$
 which are supposed to have the configuration $(uudc\overline{c})$ 
 have been estimated in the framework of the quasiparticle model
 of diquarks considering $[ud][uc]\overline{c}$ configuration. The
 masses are reproduced very well which indicates that the description
 of diquark as quasiparticle is very useful for describing multiquark
  state and to understand the dynamics of it.}}

\vskip .5in

Recently LHCb[1] has reported the existence of pentaquark charmonium states
with the decay of $\Lambda_{b}^{0}$ $(\Lambda_{b}^{0} \rightarrow J\psi K^{-}p)$.
The intermediate states have been identified as $P_{c}^{*}(4380)$ and $P_{c}^{*}(4450)$.
The states are identified as sum of two up quarks,
one down quark, one charm quark and one anti-charm quark with 
spin $\frac{3}{2}$ and $\frac{5}{2}$ respectively.
The identification of these pentaquark states is exciting and will 
give new impetus to the study of the properties and dynamics of 
multiquark states [2]. The pentaquarks are usually 
described as diquark-diquark-antiquark configuration [3,4]. The diquarks which 
are supposed to be highly co-related antisymmetric coloured object are 
one of the important candidate for study of the exotic particles as well as usual baryons.
 A number of models have been suggested for diquark [3-10]. In the present work we have studied the recently identified 
 $P_{c}^{*}$ states in diquark-diquark-antiquark configuration in the framework of quasiparticle 
 model of diquark suggested by us [10-12].

\vskip .01in

Recently we have suggested a model for diquark in which two quarks are
assumed to be correlated to form a low energy configuration 
[10-12]. Diquarks are supposed to behave like a  quasi particle in
an analogy with an electron in the crystal lattice which behaves
as a quasi particle [13]. It is well known that a quasi particle
is a low-lying excited state whose motion is modified by the
interactions within the system. An electron in a crystal is
subjected to two types of forces one is the effect of the
crystal field (-$\nabla V$) and the other is external force (F) which
accelerates the electron [14]. Under the influence of these two
forces, it behaves like a quasi particle
having velocity $v$ whose effective mass $m^{*}$ reflects the
inertia of electrons in a crystal field and can be represented as [13]:

\begin {equation}
m^{*}\frac{dv}{dt}= F
\end{equation}

The bare electrons (with normal mass m) are affected by the lattice
force -$\nabla V$ and the external
force F so that:

\begin{equation}
m\frac{dv}{dt}= F - \frac{dV}{dx}
\end{equation}

 From (1) and (2) the ratio of the normal mass (m) to the effective mass ($m^*$) can be
represented as:

 \begin{equation}
 m/m^{*}= 1 - \frac{1}{F}[\frac{\delta{V}}{\delta x}]
 \end{equation}

The difference between the effective and normal mass is attributed to the
lattice force. The sign of its average $m^{*}$
can be less or greater than 'm' or even negative according to the sign of the potential.
An elementary particle in vacuum may be suggested to be in a
situation exactly resembling that of an electron in a crystal
[14]. We have proposed a similar type of picture for the diquark
as a quasi particle inside a hadron. We have assumed that under the combined force of confinement and asymptotic 
freedom a diquark in hadron behaves like a quasi particle and 
its mass gets modified simulating the many body interaction in a hadron. The potential V
=$\frac{2}{3}$$\frac{\alpha_{s}}{r}$ (where $\alpha_{s}$ is the
strong coupling constant) is assumed to resemble
the crystal field on a crystal electron and is positive for resonance state [13]. On the other hand 
 an average force F = -ar resembles the external force where
'a' is a suitable constant. The potential can be represented as:

\begin{equation}
V_{ij}=\frac{\alpha}{r}+(F_{i}.F_{j})(-\frac{1}{2}Kr^{2})
\end{equation}

Where the coupling constant
$\alpha$=(2/3)$\alpha_{s}$,$F_{i}$.$F_{j}$=-(2/3), K is the
strength parameter. Hence $V_{ij}$ may be represented as:

\begin{equation}
V_{ij}=\frac{(2/3)\alpha_{s}}{r}+ar^{2}
\end{equation}
Where a=K/3.

The ratio of the constituent mass and the effective mass of the
diquark (m$_{D}$) has been obtained following equation (3) as,

 \begin{equation}
 \frac{m_{q}+m_{q\prime}}{m_{D}}= 1- \frac{\alpha_{s}}{3ar^{3}}
 \end{equation}
 
here $m_{q}+m_{q\prime}$ represents the normal constituent mass of
the diquark, $m_{D}$ is the effective mass of the diquark and 'r' is radius parameter of diquark.
 With $\alpha_{s}$= 0.2 $[15]$, a = 0.02 $GeV^{3}$[14], $r_{ud}$(scalar)= 0.98fm [5], $r_{ud}$
(vector)= 0.8fm[16], $r_{uc}$(scalar)= 1.1fm[17],$r_{uc}$(vector) =
0.861fm [5], $m_{u}$=$m_{d}$= 0.360 GeV [4] and $m_{c}$= 1.55 GeV [18] 
we have estimated the masses of the diquarks in the framework of the quasi
particle model [14]. We have obtained the diquark mass values as $m_{ud}$ = 0.763 GeV, $m_{uc}$ = 1.989 GeV
for scalar diquarks and $m_{ud}$= 0.803 GeV, $m_{uc}$ = 2.084 GeV for vector diquark.
It is interesting to observe here that simulating the many body
interaction the effective mass of the diquark becomes greater than the
constituents.

\vskip .05in

The mass formula of the pentaquark state can be expressed with relevant diquark-diquark-antiquark configuration as,

\begin{equation}
M = m_{D_{1}} + m_{D_{2}} + m_{\overline{q}} + E_{S}
\end{equation}

where $m_{D_{1}}$, $m_{D_{2}}$ are diquark masses, $m_{\overline{q}}$ is the antiquark mass and
 $E_{S}$ is spin term and expressed as [19,20]:

\begin{equation}
E_{S}= \frac{8}{9} \frac{\alpha_{S}}{m_{1}m_{2}}
\vec{S}_{1}\cdot\vec{S}_{2} |\psi(0)|^{2}
\end{equation}

where the strong interaction constant $\alpha_{S}$ = 0.2 [15] and
$\vec{S}_{1}\cdot\vec{S}_{2}$ is the spin interaction of corresponding states.
The masses have been estimated using the relation (7) and displayed in Table I.

\vskip 0.1in

\begin{center}

Table I:  Mass of pentaquark charmonium states:
\vskip 0.1in
\begin{tabular}{r r r r }
\hline
\hline
$Pentaquark$&$Quark $&$Experimental$&$Our$\\
$charmonium$&$Content$&$Mass^{[1]}$&$Work$\\
$state$&$$&$(MeV)$&$(MeV)$\\

\hline
$P_{c}^{*}(4380) $&$[ud]_{0}[uc]_{1}\overline{c}$&$4380\pm 8$&$ 4400 $\\
$(spin \frac{3}{2}) $&$[ud]_{1}[uc]_{0}\overline{c}$&$\pm 29$&$ 4345 $\\
$ $&$ $&$ $&$ $\\

$P_{c}^{*}(4450)  $&$ [ud]_{1}[uc]_{1}\overline{c}$ & $ 4449.8\pm 1.7$ & $ 4443$\\
$(spin \frac{5}{2})  $&$ $&$ \pm 2.5$&$ $\\

\hline
\hline
\end{tabular}

\end{center}

In the present work we have estimated masses of the particles
$P_{c}^{*}(4380)$of spin $\frac{3}{2}$ in both
$[ud]_{0}[uc]_{1}\overline{c}$ and $[ud]_{1}[uc]_{0}\overline{c}$
configurations and have obtained the masses
as 4403 MeV and 4345 MeV respectively. The $P_{c}^{*}(4450)$ of spin $\frac{5}{2}$
has been estimated in the pentaquark configuration of
$ [ud]_{1}[uc]_{1}\overline{c}$ and obtained the mass as 4443 MeV.
The results are found to be in very good agreement with experiment. In the current work 
it is observed that the description of pentaquark as diquark-diquark-antiquark
picture with diquark as quasi particle reproduces the observed mass of intermediate state very well.
It is interesting to observe that the contribution from vacuum can be
simulated as effective mass approximation for diquark
whose effective mass is more than the constituents and in excited state. 
The experimental identification of pentaquark is long awaited. 
In our present investigation the mass of $P_{c}^{*}(4380)$ and $P_{c}^{*}(4450)$ have been described
in quasiparticle picture of 'diquark'. We will also study the particles as 
baryon-meson system and also in composite fermion model of diquark in our 
future work. The current investigation has immense importance in the understanding of 
quark dynamics in multiquark system and found to be very useful and prospective.   

\vskip .1in

 {\bf Acknowledgments }:

Authors are thankful to University Grants Commission, New Delhi,
India for their financial supports.

\pagebreak {\bf References}:-

\noindent [1] R. Aaij, B. Adeva, M. Adinolfi, A. Affolder, Z. Ajaltouni,
S. Akar, J. Albrecht, F. Alessio, M. Alexander, S. Ali et al, (LHCb Collaboration), arXiv:1507.03414v1[hep-ex] (2015).

\noindent [2] M. Gell-Mann, Phys. Lett. {\bf 8}, 214 (1964).

\noindent [3] R. L. Jaffe and F. Wilczek, Phys. Rev. Lett. {\bf 91}, 232003 (2003).

\noindent [4] M. Karlinar and H. Lipkin, Phys. Lett. B {\bf 575}, 294 (2003).

\noindent [5]  A. S. de Castro, H. F. de Carvalho and A. C. B. Antunes, Z. Phys. C {\bf 57}, 315 (1993).

\noindent [6] M. Anselmino, E. Predazzi, S. Ekelin, S. Fredriksson and D. B. Lichenberg, Rev. Mod. Phys. {\bf 65}, 1199 (1993).

\noindent [7] T. Schafer and E. Shuryak, Rev. Mod. Phys. {\bf 70}, 323 (1998).

\noindent [8] R. G. Betman and L. V. Laparashvili, Sov. J. Nucl. Phys. {\bf 41}, 295 (1985).

\noindent [9] E. V. Shuryak, Nucl. Phys. B {\bf 203}, 93 (1982);
              E. V. Shuryak and I. Zahed, Phys. Lett. {\bf 589}, 21 (2004).

\noindent [10] A. Bhattacharya, A. Sagari, B. Chakrabarti and S. Mani, Phys. Rev. C {\bf 81}, 015202 (2010).

\noindent [11] A. Bhattacharya, B. Chakrabarti, A. Sagari and S. Mani, Int. J. Theo. Phys. {\bf 47}, 2507 (2008).

\noindent [12] A. Bhattacharya, A. Chandra, B. Chakrabarti and A. Sagari, Eur. Phys. J. Plus. {\bf 126}, 57 (2011).

\noindent [13] A. Haug, $\emph{Theoretical Solid State Physics}$ (Pergamon Press,1975),Vol.1, p.100  .

\noindent [14] B. Chakrabarti, A. Bhattacharya, S. Mani and A. Sagari, Acta. Phys. Polo. B{\bf 41} 95 (2010).

\noindent [15] W. Lucha, F. F. Scholberl and d. Gromes, Phys. Rep. {\bf 200}, 168 (1991).

\noindent [16] K. Nagata and A. Hosaka, Ann. Rep./2006/Sec 2/nagata.pdf.

\noindent [17] B. Chakrabarti, Mod. Phys. Lett. A {\bf 12}, 2133 (1997).

\noindent [18]  Griffiths, David, $\emph{Introduction to Elementary Particles}$ (WILEY-VCH, 2008) p.135.

\noindent [19]. A.Bhattacharya, B. Chakrabarti, T. Sarkhel and S. N. Banerjee, Int. J. Mod. Phys. A {\bf 15}, 2053 (2000).

\noindent [20]. A. Bhattacharya, B. Chakrabarti and S. N. Banerjee, Eur. Phys. J. C {\bf 2}, 671 (1998).

\end{document}